\documentclass{pasj01}
\usepackage{natbib} 
\usepackage{lscape}
\usepackage{url}
\Received{$\langle$reception date$\rangle$}
\Accepted{$\langle$acception date$\rangle$}
\Published{$\langle$publication date$\rangle$}

\newcommand{\oi}{[O{\sc i}]}
\newcommand{\oiii}{[O{\sc iii}]}

\newcommand{\cii}{[C{\sc ii}]}

\newcommand{\nii}{[N{\sc ii}]}

\newcommand{\namea}{J2054-0005}
\newcommand{\nameb}{J2310+1855}

\newcommand{\lsun}{$L_{\rm \odot}$}
\newcommand{\msun}{$M_{\rm \odot}$}
\newcommand{\ltir}{$L_{\rm TIR}$}
\newcommand{\td}{$T_{\rm d}$}

\newcommand{\lcii}{$L_{\rm [CII]}$}

\newcommand{\pasa}{PASA}

\begin{document}

\title{
Detections of \oiii\ $88$ \micron\ in Two Quasars in the Reionization Epoch
}
\author{
Takuya Hashimoto\altaffilmark{1,2,3}, 
Akio K. Inoue\altaffilmark{1,2,4,5}, 
Yoichi Tamura\altaffilmark{6},
Hiroshi Matsuo\altaffilmark{3,7}, 
Ken Mawatari\altaffilmark{8}, 
and 
Yuki Yamaguchi\altaffilmark{9}
}
\altaffiltext{1}{Faculty of Science and Engineering, Waseda University, 3-4-1 Okubo, Shinjuku, Tokyo 169-8555, Japan}
\altaffiltext{2}{Department of Environmental Science and Technology, Faculty
of Design Technology, Osaka Sangyo University, 3-1-1, Nagaito,
Daito, Osaka 574-8530, Japan}
\altaffiltext{3}{National Astronomical Observatory of Japan, 2-21-1 Osawa, Mitaka, Tokyo 181-8588, Japan}
\altaffiltext{4}{Department of Physics, School of Advanced Science and Engineering, Faculty of Science and Engineering, Waseda University, 3-4-1 Okubo, Shinjuku, Tokyo 169-8555, Japan}
\altaffiltext{5}{Research Institute for Science and Engineering, Waseda University, 3-4-1 Okubo, Shinjuku, Tokyo 169-8555, Japan}
\altaffiltext{6}{Division of Particle and Astrophysical Science, Graduate School of Science, Nagoya
University,  Furo-cho, Chikusa-ku, Nagoya, Aichi 464-8602, Japan} 
\altaffiltext{7}{Department of Astronomical Science, School of Physical Sciences, The Graduate University for Advanced Studies (SOKENDAI),2-21-1, Osawa, Mitaka, Tokyo 181-8588, Japan}
\altaffiltext{8}{Institute for Cosmic Ray Research, The University of Tokyo, Kashiwa, Chiba 277-8582, Japan}
\altaffiltext{9}{The University of Tokyo, 2-21-1 Osawa, Mitaka, Tokyo 181-0015, Japan}
\email{
thashimoto@obsap.phys.waseda.ac.jp
}

\maketitle

\begin{abstract}
With the Atacama Large Millimeter/sub-millimeter Array (ALMA), we report detections of the far-infrared (FIR) \oiii\ 88 \micron\ line and the underlying dust continuum in the two quasars in the reionization epoch, J205406.48-000514.8 (hereafter \namea) at $z=6.0391\pm0.0002$ and J231038.88+185519.7 (hereafter \nameb) at $z=6.0035\pm0.0007$. 
The \oiii\ luminosity of \namea\ and \nameb\ are $L_{\rm [OIII]} = (6.8\pm0.6) \times 10^{9}$ and $(2.4\pm0.6) \times 10^{9}$ \lsun, corresponding to $\approx$ 0.05\%\ and 0.01\%\ of the total infrared luminosity, \ltir, respectively. Combining these \oiii\ luminosities with \cii\ 158 \micron\ luminosities in the literature, we find that \namea\ and \nameb\ have the \oiii-to-\cii\ luminosity ratio of $2.1\pm0.4$ and $0.3\pm0.1$, respectively, the latter of which is the lowest among objects so far reported at $z>6$. 
Combining \oiii\ observations at $z\approx6-9$ from our study and the literature, we identify the \oiii\ line deficit: objects with larger \ltir\ have lower $L_{\rm [OIII]}$-to-\ltir\ ratios. Furthermore, we also find that the anti-correlation is shifted toward higher \ltir\ value when compared to the local \oiii\ line deficit. 
\end{abstract}

\section{Introduction}
\label{sec:intro}

Quasars are powered by supermassive black holes (SMBHs) with $\approx 10^{8-10}$ \msun\ (e.g., \citealt{de.rosa2014, wu2015}). Owing to wide-area surveys, $\approx 100$ quasars are discovered at $z>6$ (e.g., \citealt{fan2003, jiang2016, mazzucchelli2017, matsuoka2018}) and up to $z=7.54$ (\citealt{banados2018}). How SMBHs have accreted within $\approx1$ Gyr after the Big Bang is one of the most important question in modern astronomy (\citealt{valiante2017}). 

In the local Universe, there is a tight correlation between the central black hole mass and the bulge mass (\citealt{haering2004, kormendy.ho2013}). Given the coevolution of the SMBHs and their host galaxies, understanding the host galaxy properties at the earliest Universe is crucial.
Rest-frame far-infrared (FIR) dust continuum observations show that high-$z$ quasar host galaxies have star formation rates (SFRs) $\approx50-2700$ \msun\ yr$^{-1}$ and large dust masses $\approx 10^{7}-10^{9}$ \msun\ (e.g., \citealt{wang2008, venemans2018}). The carbon monoxide (CO) line observations reveal a large amount of gas mass in the host galaxies ($\approx10^{10}$ \msun) (e.g., \citealt{wang2010, venemans2017.co, feruglio2018}). 
The FIR fine structure line of \cii\ 158 \micron\ is widely used to obtain the precise redshift and  the dynamical mass (e.g., \citealt{maiolino2005, wang2013, venemans2017.z7.5, decarli2018, izumi2018}). 

Combinations of multiple FIR fine structure lines are useful to obtain physical properties of the interstellar medium (ISM) such as the gas-phase metallicity, the electron density, and the ionization parameter (e.g., \citealt{nagao2011, pereira-sataella2017}). Among the FIR lines, the \oiii\ 88.356 \micron\ line ($\nu_{\rm rest}=$ 3393.006244 GHz) would be a good next target after \cii\ because it is the second most commonly observed line in normal star-forming galaxies at $z>6$ (e.g., \citealt{inoue2016, tamura2019}). Indeed, recent ALMA observations demonstrate that \oiii\ is detectable even at $z=9.11$ (\citealt{hashimoto2018}). 

In this paper, we report results of our ALMA Band 8 observations targeting \oiii\ in two quasars at $z\approx6$, J205406.48-000514.8 (hereafter \namea) and J231038.88+185519.7 (\nameb). With our observations (\S \ref{sec:sample_data}), we successfully detect \oiii\ and the underlying dust continuum (\S \ref{sec:result}). In conjunction with \cii\ measurements in the literature, we discuss their [OIII]-to-[CII] line luminosity ratios (\S \ref{sec:discussion}). 
Throughout this paper, we adopt a flat $\Lambda$CDM cosmology ($\Omega_{\small m} = 0.272$, $\Omega_{\small \Lambda} = 0.728$, and $H_{\small 0} = 70.4$ km s$^{-1}$ Mpc$^{-1}$; \citealt{komatsu2011}). The solar luminosity, \lsun, is $3.839\times10^{33}$ erg s$^{-1}$, and $k_{\rm B}$ represents the Boltzmann constant.

\section{Our Sample and ALMA Band 8 Data}
\label{sec:sample_data}

\subsection{Sample}
\label{subsec:sample}

At the time of writing our proposal, April 2017, there were 13 quasars with \cii\ detections at $z\geq6.0$\footnote{J1148+5251 (\citealt{maiolino2005, maiolino2012}), J112+0641 (\citealt{venemans2012}), J2348-3054, J0109-3047, J0305-3150 (\citealt{venemans2016}), J2310+1855, J1319+0950, J2054-0005 (\citealt{wang2013}), J0100+2802 (\citealt{wang2016}), P036+03 (\citealt{banados2015}), J0210-0456 (\citealt{willott2013}), J0055+0146, and J2229+1457 (\citealt{willott2015.qso}).}.
We excluded an object with declination too high for ALMA observations. We then omitted four objects with redshifts at which \oiii\ emission is strongly affected by atmospheric absorption\footnote{J1148+5251 is excluded from the candidates because of its high declination for ALMA observations. Four objects, J0305-3150, J1319+0950, P036+03, and J2229+1457 are additionally omitted because their \oiii\ frequencies are strongly affected by atmospheric absorption.}.
To secure the \oiii\ line detection within reasonable ALMA integration times, we selected objects with (i) bright total infrared luminosities, \ltir, and (ii) relatively lower-$z$ among the candidates. Finally, these leave us with two objects, \namea\ and \nameb, which have very bright total infrared luminosities, log(\ltir/\lsun) $\approx13$, at $z=6.0$. In fact, \nameb\ has the brightest infrared, \cii\ and CO(6-5) luminosities at $z\geq6.0$ (\citealt{decarli2018, venemans2018, feruglio2018}).

These objects are originally discovered by the  Sloan Digital Sky Survey data (\citealt{jiang2008, jiang2016}). \namea\ (\nameb) has the UV absolute magnitude of $M_{\rm 1450} =-26.1$ ($-27.8$) and the bolometric luminosity of $2.8\times10^{13}$ \lsun\ ($9.3\times10^{13}$ \lsun) (\citealt{wang2013}). The BH mass in \namea\ (\nameb) is estimated to be $0.9^{+1.6}_{-0.6}\times10^{9}$ \msun\ ($2.3^{+5.1}_{-1.8}\times10^{9}$ \msun) under the assumption of the Eddington-limited mass accretion (\citealt{wang2013, willott2015.qso}). The \cii\ redshift value of \namea\ (\nameb) is  $z=6.0391\pm0.0002$ ($6.0031\pm0.0002$) (\citealt{wang2013}). 

\subsection{Observations and Data}
\label{subsec:data}

We performed observations of \oiii\ with ALMA Band 8 during 2018 March and 2018 July (ID 2017.1.01195.S, PI: T. Hashimoto). In \namea\ (\nameb), 43 antennas with the baseline lengths of $15 - 785$ m ($15-360$ m) were used, and the total on-source exposure time was 127 minutes (176 minutes). 
Four spectral windows (SPWs) with a bandwidth of 1.875 GHz were used in the Frequency Division Mode. Two slightly overlapping SPWs (0 \&\ 1) were used to target \oiii, covering the frequency range of $480.71-483.68$ GHz ($483.19-486.31$ GHz) for \namea\ (\nameb). The other two SPWs (2 \&\ 3) were used to observe the continuum, covering $492.21-495.96$ GHz ($494.70-498.45$ GHz) for \namea\ (\nameb). 
A quasar J1924-2914 (J2258-2758) was used for bandpass and flux calibrations, and a quasar J2101+0341 (J2253+1608) was used for phase calibrations.The data were reduced and calibrated using CASA pipeline version 5.1.1-5. We produced images and cubes with the {\tt CLEAN} task using the natural weighting.  To create a pure dust continuum image, we collapsed all off-line channels. To create a pure line image, we subtracted continuum using the off-line channels in the line cube with the CASA task {\tt uvcontsub}. In \nameb, we could not obtain the data product in SPW1 due to very strong atmospheric absorption\footnote{According to the QA2 Report, ALMA staffs tried to keep a part of the data by changing parameters of the pipeline, but it did not work.}. 

With the CASA task {\tt imstat},we estimate the rms level of the continuum image of \namea\ (\nameb) to be 67 $\mu$Jy beam$^{-1}$ (106 $\mu$Jy beam$^{-1}$). The spatial resolution of the continuum image is $0''.38 \times 0''.34$ ($0''.69 \times 0''.60$) in FWHM with a beam position angle, PA, of 69$^{\circ}$ ($-61^{\circ}$). The typical rms level of the line cube is $0.6$ mJy beam$^{-1}$ (0.8 mJy beam$^{-1}$) per 30 km s$^{-1}$ bin. 

\section{Results}
\label{sec:result}

\begin{table*}[t]
\tbl{Summary of observational results. \label{tab1}}
{
\begin{tabular}{lcc}
\hline
& \namea & \nameb \\
\hline
$z_{\rm [OIII]}$ & $6.0391\pm0.0002$ & $6.0035\pm0.0007$ \\
FWHM(\oiii) [km s$^{-1}$] & $282\pm17$ & $333\pm72$ \\ 
\oiii\ integrated flux [Jy km s$^{-1}$]& $3.79\pm0.34$ & $1.38\pm0.34$ \\ 
\oiii\ luminosity [$10^{9}$ \lsun] & $6.79\pm0.61$ & $2.44\pm0.61$ \\ 
$S_{\rm \nu,87}$ [mJy] & 10.35 $\pm$ 0.15 & 24.89 $\pm$ 0.21 \\
Dust deconvolved size$^{a}$ [arcsec$^{2}$] & (0.23 $\pm$ 0.01) $\times$ (0.15 $\pm$ 0.02) & (0.31 $\pm$ 0.01) $\times$ (0.22 $\pm$ 0.02) \\ 
\oiii\ deconvolved size$^{a}$  [arcsec$^{2}$] & $(0.49\pm0.07) \times (0.45\pm0.06)$ & $(0.44\pm0.27) \times (0.38\pm0.13)$\\
ALMA dust position (ICRS) & $20^h 54^m 06\fs 503, -00\arcdeg 05' 14\farcs43$ & $23^h 10^m 38\fs 902, +18\arcdeg 55' 19\farcs83$\\
ALMA \oiii\ position (ICRS) & $20^h 54^m 06\fs 503, -00\arcdeg 05' 14\farcs48$ & $23^h 10^m 38\fs 900, +18\arcdeg 55' 19\farcs80$\\
SDSS optical position$^{b}$ (ICRS) & $20^h 54^m 06\fs 486, -00\arcdeg 05' 14\farcs50$ & $23^h 10^m 38\fs 882, +18\arcdeg 55' 19\farcs61$\\
\td\  [K] & $50\pm2$ & $37\pm1$ \\
$\beta_{\rm d}$ & $1.8\pm0.1$ & $2.2\pm0.1$ \\
\ltir$^{c}$  [$10^{13}$ \lsun] & $1.3^{+0.2}_{-0.2}$ & $1.9^{+0.2}_{-0.1}$ \\
SFR$_{\rm IR}$$^{d}$ [\msun\ yr$^{-1}$] & $1897^{+265}_{-216}$ & $2873^{+294}_{-232}$ \\
\hline
\end{tabular}
}
\tabnote{Note. 
$^{a}$ The values represent major and minor-axis FWHM values of a 2D Gaussian profile. 
$^{b}$ The SDSS optical position recalibrated with the GAIA's astrometry (see \S \ref{subsec:astrometry}).
$^{c}$ The total luminosity, \ltir, is estimated by integrating the modified-black body radiation at $8-1000$ \micron.
$^{d}$ The SFR value is obtained following \cite{kennicutt.evans2012} under the assumption of the Kroupa initial mass function (IMF; \citealt{kroupa2001}) in the range of $0.1-100$ \msun. The value should be treated as the upper limit on the obscured SFR (see \S \ref{subsec:discussion1}).
}
\end{table*}

\begin{figure*}[hbtp]
\hspace{+2.5cm}
\vspace{+0.5cm}
\includegraphics[width=12cm]{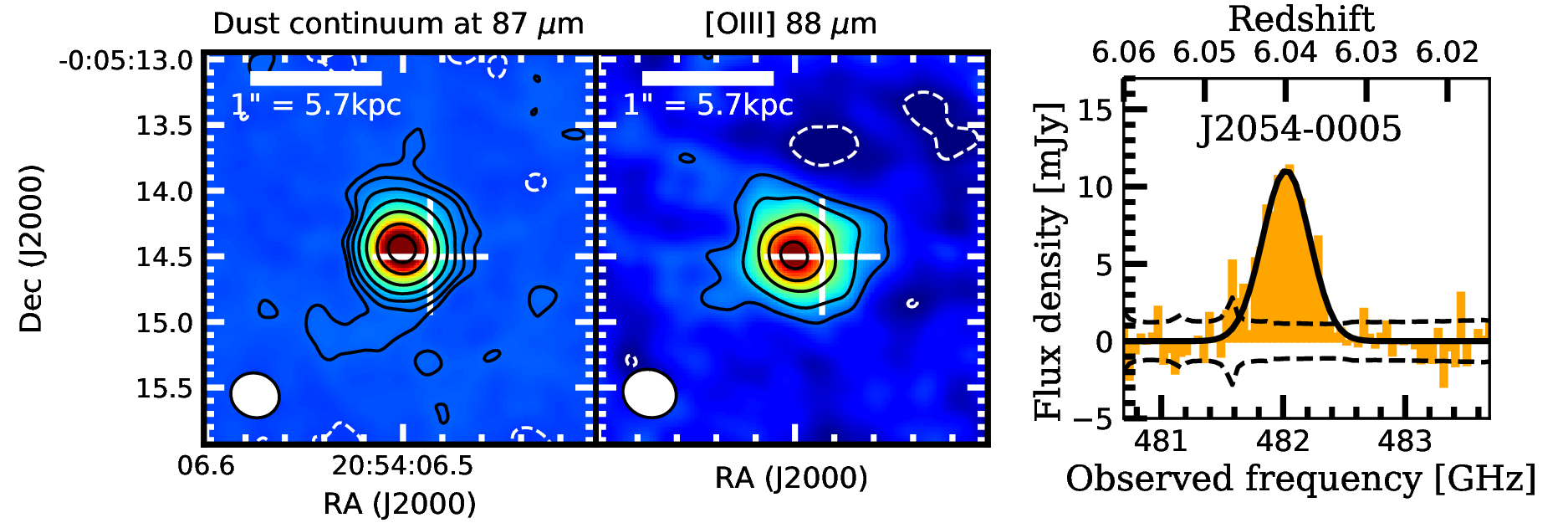}

\hspace{+2.5cm}
\vspace{+0.5cm}
\includegraphics[width=12cm]{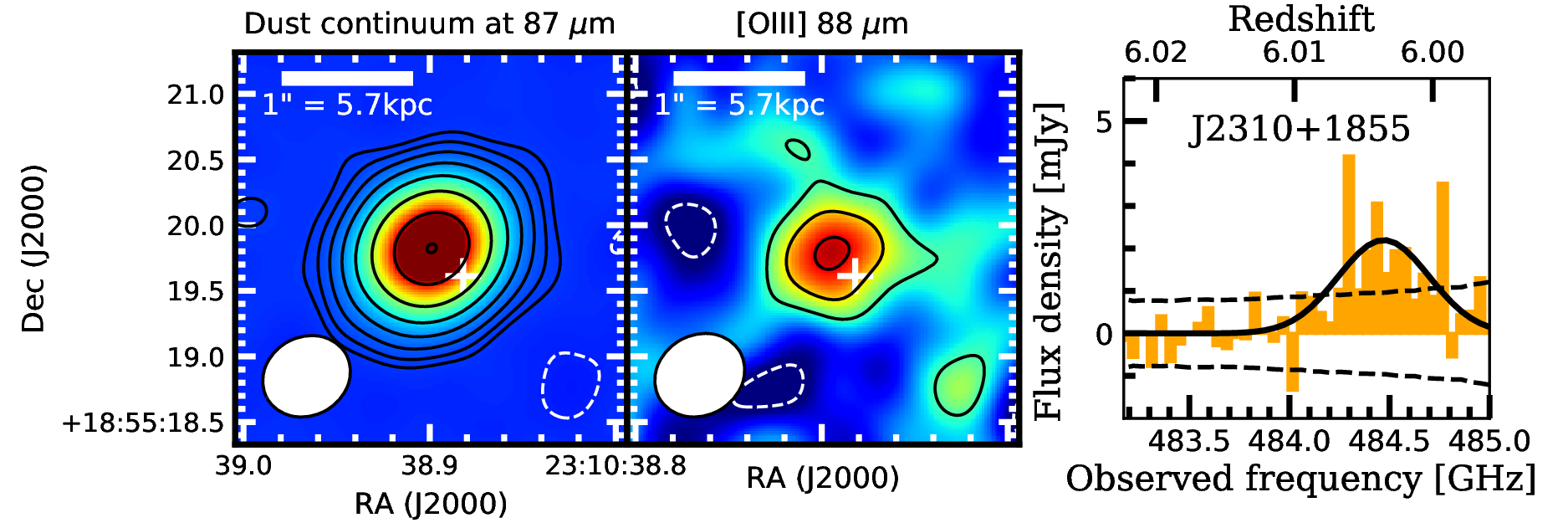}
\caption
{
The dust continuum image at $\approx 87$ \micron\ (left), the \oiii\ 88 \micron\ line image (middle), and the continuum-subtracted \oiii\ spectrum (right). 
In the left and middle panels, the ellipse at lower left corner indicates the synthesized beam size of ALMA, and the scale bar is shown at the upper left corner. Negative and positive contours are shown by the white dashed and black solid lines, respectively. The white crosses show the optical position in the SDSS $z$-band image whose astrometry is calibrated by the GAIA astrometry (see \S \ref{subsec:oiii}). The size of the white cross corresponds to the $1\sigma$ positional uncertainty of the optical counterpart. We do not see significant ($>3\sigma$) positional offsets between the ALMA and optical images. In the right panel, the continuum-subtracted \oiii\ spectrum is extracted from the region with $>3\sigma$ detections in the velocity-integrated intensity images. The black solid lines are the best-fit Gaussian for the \oiii\ line, while the black dashed lines show the $\pm1\sigma$ noise levels. 
{\bf \namea\ (Top)}-- The dust continuum contours drawn at (-2, 2, 4, 8, 16, 32, 64, 100) $\times \sigma$, where $\sigma =$ 67 $\mu$Jy beam$^{-1}$. The \oiii\ line contours drawn at (-2, 2, 4, 8, 12, 16) $\times \sigma$, where $\sigma =$ 94 mJy beam$^{-1}$ km s$^{-1}$. 
{\bf \nameb\ (Bottom)}-- The dust continuum contours at $\approx 87$ $\mu$m drawn at (-2, 2, 4, 8, 16, 32, 64, 128, 200) $\times \sigma$, where $\sigma =$ 106 $\mu$Jy beam$^{-1}$. The \oiii\ line contours drawn at (-2, 2, 4, 6) $\times \sigma$, where $\sigma =$ 149 mJy beam$^{-1}$ km s$^{-1}$. 
}
\label{fig1}
\end{figure*}

\subsection{Dust Continuum}
\label{subsec:dust}

Our data probe dust continuum emission at the rest-frame wavelength, $\lambda_{\rm rest}$, of  $\approx 87$  \micron. The top left and bottom left panels of Figure \ref{fig1} show dust continuum images of \namea\ and \nameb, respectively. Our measurements are summarized in Table \ref{tab1}. 

{\bf \namea}-- 
To estimate the flux density and the beam deconvolved size of the dust continuum, we apply the CASA task {\tt imfit} assuming a 2D Gaussian profile for the specific intensity. We estimate the continuum flux density to be $S_{\nu, 87\mu m} =$ 10.35 $\pm$ 0.15 mJy. The beam deconvolved size is (0.23 $\pm$ 0.01) $\times$ (0.15 $\pm$ 0.02)  arcsec$^{2}$, corresponding to (1.34 $\pm$ 0.06) $\times$ (0.88 $\pm$ 0.13) kpc$^{2}$ at $z=6.0391$, with PA $=$177$^{\circ}$ $\pm$ 7$^{\circ}$.

{\bf \nameb}--
The continuum flux density is 24.89 $\pm$ 0.21 mJy. The beam deconvolved size is (0.31 $\pm$ 0.01) $\times$ (0.22 $\pm$ 0.02)  arcsec$^{2}$, corresponding to (1.81 $\pm$ 0.06) $\times$ (1.28 $\pm$ 0.13) kpc$^{2}$ at $z=6.0391$, with PA $=$ 154$^{\circ}$ $\pm$ 8$^{\circ}$. 

These deconvolved size and PA values are consistent with those obtained by \cite{wang2013} using ALMA Band 6 data within $1-2\sigma$ uncertainties\footnote{\cite{wang2013} have used the CASA task {\tt imfit} to obtain the beam-deconvolved sizes of \cii\ and dust continuum emitting regions in the same manner as used in this study.}. 


\subsection{\oiii\ 88 \micron}
\label{subsec:oiii}

The \oiii\ is detected in the two quasars. 
Our measurements are summarized in Table \ref{tab1}. 
The top (bottom) middle panel of Figure \ref{fig1} shows a velocity-integrated intensity image between $481.7 - 482.6$ GHz ($484.1 - 485.0$ GHz) for \namea\ (\nameb). The peak intensity is $1.67\pm0.10$ Jy km s$^{-1}$ beam$^{-1}$ ($0.94\pm0.15$ Jy km s$^{-1}$ beam$^{-1}$). We perform photometry on the image with the CASA task {\tt imfit} assuming a 2D Gaussian profile for the line intensity. 

In \namea, the total line flux is estimated to be $3.79\pm0.34$ Jy km s$^{-1}$. The beam-deconvolved size is $(0.49\pm0.07) \times (0.45\pm0.06)$ arcsec$^{2}$, corresponding to  $(2.87\pm0.41) \times (2.63\pm0.35)$ kpc$^{2}$ at  $z=6.0391$, with PA $=75^{\circ}\pm82^{\circ}$. 
Likewise, in \nameb, the total line flux is estimated to be $1.38\pm0.34$ Jy km s$^{-1}$. The beam-deconvolved size is $(0.44\pm0.27) \times (0.38\pm0.13)$ arcsec$^{2}$, corresponding to $(2.57\pm1.58) \times (2.22\pm0.76)$ kpc$^{2}$ at $z=6.0035$, with PA $=70^{\circ}\pm97^{\circ}$. 
We note that the two quasars have the \oiii\ emitting region size of $\approx2-3$ kpc (FWHM), which is significantly larger than the continuum emitting region size of $\approx 1$ kpc (FWHM).

The top (bottom) right panel of Figure \ref{fig1} shows the continuum-subtracted spectrum of \namea\ (\nameb) extracted from the \oiii\ region with $>3\sigma$ detections in the velocity-integrated intensity image. We obtain the \oiii\ redshift of  $6.0391\pm0.0002$ ($6.0035\pm0.0007$) and the FWHM value of $282\pm17$ km s$^{-1}$ ($333\pm72$ km s$^{-1}$). 
Based on a combination of the flux and redshift values, we obtain the \oiii\ luminosities of (6.79 $\pm$ 0.61) $\times 10^{9}$ and  (2.44 $\pm$ 0.61) $\times 10^{9}$ \lsun\ in \namea\ and \nameb, respectively. 

To investigate a possible broad velocity component in the \oiii\ line, as that found in a $z=6.4$ quasar in \cii\  (\citealt{maiolino2012}), we extract two additional spectra from the \oiii\ regions with $>1\sigma$ and $>2\sigma$ detections in the velocity-integrated intensity images. We do not find any broad velocity component in the spectra.

We compare our \oiii\ measurements with \cii\ measurements presented in \cite{wang2013}. In \namea, the \oiii\ emitting region size, $(0.49 \pm 0.07) \times (0.45 \pm 0.06)$ arcsec$^{2}$, is consistent with the \cii\ emitting region, $(0.35\pm0.04) \times (0.32\pm0.05)$ arcsec$^{2}$, within $\approx2\sigma$ uncertainties. Likewise, the \oiii\ line FWHM, $282 \pm 17$ km s$^{-1}$, is consistent with that of \cii, $243 \pm 10$ km s$^{-1}$, within $\approx2\sigma$ uncertainties. 
In \nameb, \oiii\ emitting region size, $(0.44 \pm 0.27) \times (0.38 \pm 0.13)$ arcsec$^{2}$, is consistent with that of \cii,  $(0.56\pm0.03) \times (0.39\pm0.04)$ arcsec$^{2}$, within $1\sigma$ uncertainties. Likewise, the \oiii\ line FWHM value, $333 \pm 72$ km s$^{-1}$, is consistent with that of \cii, $393 \pm 21$ km s$^{-1}$, within $1\sigma$ uncertainties. 
The case of \namea\ might reveal that the size of the \oiii\ emitting region and the line FWHM are larger than those of \cii. If this is the case, \oiii\ and \cii\ lines may trace different regions of the quasar. However, because these differences are only marginal ($\approx2\sigma$), we do not attempt to discuss this  further.


\subsection{Astrometry}
\label{subsec:astrometry}

In \namea\ (\nameb), we find that the spatial positions of dust continuum and \oiii\ are consistent within 48 (72) mas uncertainties. Hereafter, we use the dust continuum positions due to their high significance detections. Based on the IRAF task {\tt imexam}, \namea\ has $\rm (\alpha, \delta) = (20^h 54^m 06\fs 503, -00\arcdeg 05' 14\farcs43)$, and \nameb\ has $\rm (\alpha, \delta) = (23^h 10^m 38\fs 902, +18\arcdeg 55' 19\farcs83)$ in the International Celestial Reference System (ICRS), on which ALMA relies (Table \ref{tab1}).

We also compare positions of the ALMA  and the SDSS optical images (c.f., \citealt{shao2019, wang2019}). To do so, we first recalibrate astrometry of the SDSS $z$-band image (\citealt{eisenstein2011}) where the quasars are detected. Using nearby bright stars whose positions are accurately measured in the GAIA second data release (DR2) catalog in the ICRS frame (\citealt{gaia2016, gaia2018}), we have performed {\tt IRAF} tasks {\tt ccmap} and {\tt ccsetwcs} to recalibrate the astrometry of the SDSS image. 
The astrometry uncertainty (i.e., systemic uncertainty) is estimated to be $\approx 120$ and $100$ mas around \namea\ and \nameb, respectively, based on comparisons of bright star positions in the GAIA catalog and the SDSS image with recalibrated astrometry. In addition, we estimate the positional uncertainty arising from the IRAF task {\tt imexam} (i.e., measurement uncertainty) to be $\approx$ 350 (40) mas in \namea\ (\nameb)\footnote{We have performed the IRAF task {\tt imfit} with five cursor positions, the peak flux pixel and the four adjacent pixels, to obtain the position. We then adopt the standard deviation of the results as the 1$\sigma$ uncertainty due to the fitting.}. The relatively large uncertainty in \namea\ is due to the low signal-to-noise ratio in the $z$-band image. We regard  470 and 140 mas as the final positional uncertainty for \namea\ and \nameb, respectively. 

In the left panels of Figure \ref{fig1}, white crosses indicate the optical positions in the SDSS $z$-band image with recalibrated astrometry. The optical position of \namea\ is $\rm (\alpha, \delta) = (20^h 54^m 06\fs 486, -00\arcdeg 05' 14\farcs50)$, $\approx250$ mas offset from the ALMA's position. Likewise, the optical  position of \nameb\ is $\rm (\alpha, \delta) = (23^h 10^m 38\fs 882, +18\arcdeg 55' 19\farcs61)$, $\approx350$ mas offset from the ALMA's position. Given the positional uncertainties, we do not conclude that there is a significant ($>3\sigma$) offset between the two images (c.f., \citealt{shao2019, wang2019}). ALMA higher-angular resolution data and deeper and higher-angular resolution optical images would be useful to further investigate if there is a possible spatial offset between the two images.

\subsection{Tight Constraints on the Dust Temperature and the Infrared Luminosity}
\label{subsec:discussion1}

Previous studies often assume that FIR dust continuum emission of quasars at $\lambda_{\rm rest} \gtrsim 50$ \micron\ is mainly powered by star-formation activity with negligible contribution from active galactic nuclei (AGNs) (e.g., \citealt{leipski2013}). Assuming that FIR dust continuum emission is described as an optically-thin modified-black body radiation, $I_{\rm \nu} \propto \nu^{3+\beta_{\rm d}}$/(exp($h\nu/kT_{\rm d}$) $- 1$), we constrain the single dust temperature, $T_{\rm d}$, and the dust emissivity index, $\beta_{\rm d}$, of the two quasar host galaxies taking CMB effects into account (\citealt{da_cunha2013}) (Table \ref{tab1}). 

\begin{figure}[]
\hspace{-0.5cm}
\includegraphics[width=9cm]{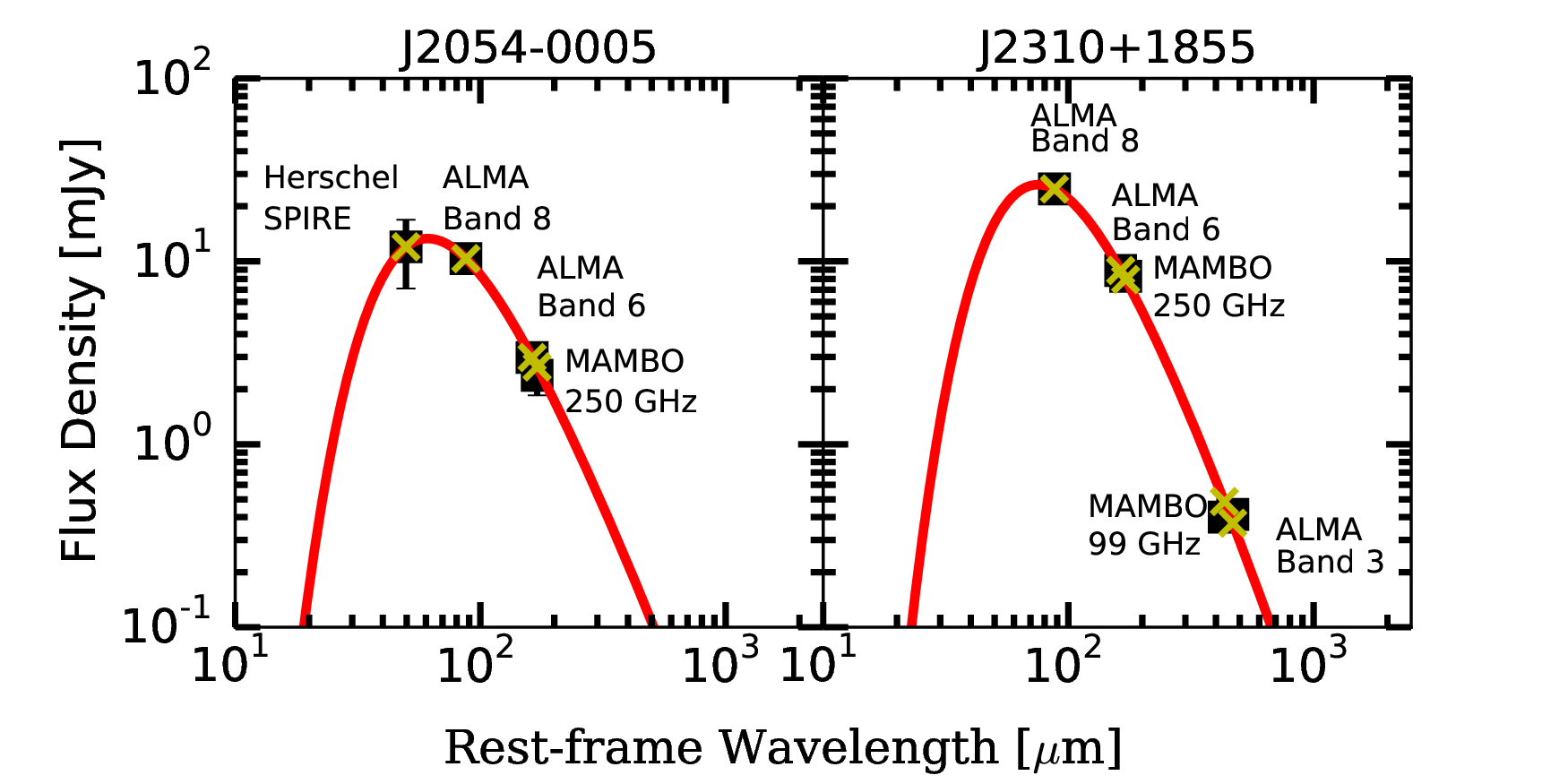}
\caption
{
Left and right panels show the FIR dust SED of \namea\ and \nameb, respectively. In each panel, black squares denote the measurements with error bars typically smaller than the symbols, while the best-fit data are shown in yellow crosses. The red line corresponds to the best-fit SED. See the text for the details of the data used in the fit. 
}
\label{fig2}
\end{figure}

In \namea, we use four flux density measurements of $12.0\pm4.9$ mJy, 10.35 $\pm$ 0.15 mJy, $2.98\pm0.05$ mJy, and $2.38\pm0.53$ mJy obtained with Herschel 350 \micron\ data (\citealt{leipski2013}), our ALMA 488 GHz, ALMA 262 GHz data (\citealt{wang2013}), and MAMBO 250 GHz data (\citealt{wang2008}), respectively. These data sample $\lambda_{\rm rest}\approx50-200$ \micron. By fitting modified-black body models corrected for the CMB effects to the photometry data, we obtain $T_{\rm d} = 50\pm2$ K and $\beta_{\rm d} = 1.8\pm0.1$ based on the $\chi^{2}$ statistics. The best-fit model is shown in the left panel of Figure \ref{fig2}. Integrating the modified-black body radiation over $8-1000$ $\mu$m, we obtain the total infrared luminosity to be \ltir\ $= 1.3^{+0.2}_{-0.2} \times 10^{13}$ \lsun. Following \cite{kennicutt.evans2012} under the assumption of the Kroupa initial mass function (IMF; \citealt{kroupa2001}) in the range of $0.1-100$ \msun, we obtain the IR-based star formation rate (SFR$_{\rm IR}$) $\approx$ 1900 \msun\ yr$^{-1}$. Note our ALMA Band 8 data are useful to constrain $T_{\rm d}$ because the data probe the wavelengths close to the peak of the dust spectral energy distribution (SED).

Likewise, in \nameb, we use five flux density measurements of 24.89 $\pm$ 0.21 mJy, $8.91\pm0.08$ mJy, $8.29\pm0.63$ mJy, $0.40\pm0.05$, and $0.41\pm0.03$ mJy obtained with our ALMA 488 GHz, ALMA 262 GHz data (\citealt{wang2013}), MAMBO 250 GHz data, MAMBO 99 GHz data (\citealt{wang2008}), and ALMA 91.5 GHz (\citealt{feruglio2018}), respectively. These data sample $\lambda_{\rm rest}\approx90-500$ \micron. We obtain $T_{\rm d} = 37\pm1$ K, $\beta_{\rm d} = 2.2\pm0.1$, \ltir\ $=1.9^{+0.2}_{-0.1} \times 10^{13}$ \lsun, and SFR$_{\rm IR} \approx 2900$ \msun\ yr$^{-1}$ in the same way as in \namea. We use these $T_{\rm d}$ and SFR$_{\rm IR}$ values to interpret our results in \S \ref{sec:discussion}.

These $T_{\rm d}$ and $\beta_{\rm d}$ values are within the ranges obtained in a mean SED of six quasar host galaxies at $z=1.8-6.4$, $T_{\rm d} = 47\pm3$ K and $\beta_{\rm d} = 1.6\pm0.1$ (\citealt{beelen2006}) and in a mean SED of seven quasar host galaxies at $z\approx4-5$,  $T_{\rm d} = 41\pm5$ K and $\beta_{\rm d} = 1.95\pm0.3$ (\citealt{Priddey.McMahon2001}). Nevertheless, our results demonstrate the variety of dust properties on the individual basis.  

In the discussion above, we have assumed that the dust continuum emission is purely powered by star formation activity. However, recent studies have decomposed the FIR SED of local AGNs and quasars into components heated by star formation and AGN activity (e.g., \citealt{symeonidis2016, ichikawa2019}); these studies show that powerful AGN activity can actually dominate the dust continuum emission up to $\lambda_{\rm rest} \approx 90$ \micron\ or longer wavelengths (see also \citealt{symeonidis2017, schneider2015}). Because the two $z\approx6$ quasars studied here are very luminous, a significant fraction of  \ltir\ could be powered by AGN activity\footnote{Recently, \cite{shao2019} have performed detailed multi-wavelength SED analyses in \nameb. Figure 4 in the study shows that star formation activity may be a dominant source for the FIR dust continuum emission in \nameb.}. 
In particular, \namea\ has a very compact dust continuum emitting region (FWHM $\approx1$ kpc), implying that its high \td\ could be largely due to heating by the quasar. 
Therefore, our values should be treated with upper limits on the dust-obscured SFRs.

\subsection{Luminosity Ratios}
\label{subsec:lumi_ratio}

\subsubsection{\oiii-Line Deficit and its Redshift Evolution}
\label{subsubsec:lumi_ratio1}

\begin{figure*}[]
\hspace{+1.5cm}
\includegraphics[width=5.5cm, angle = 270]{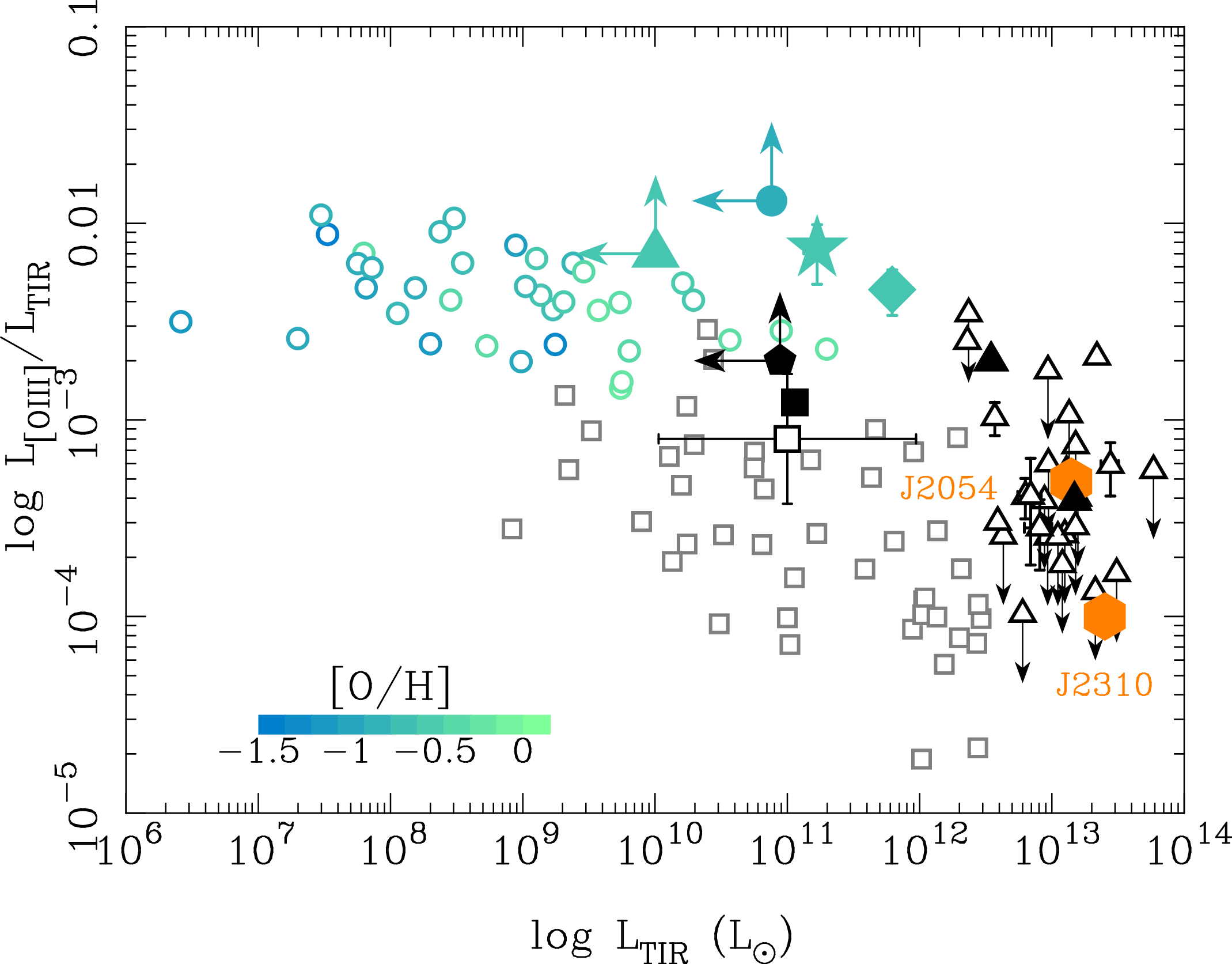}

\vspace{-5.7cm}
\hspace{+9.5cm}
\includegraphics[width=6.0cm]{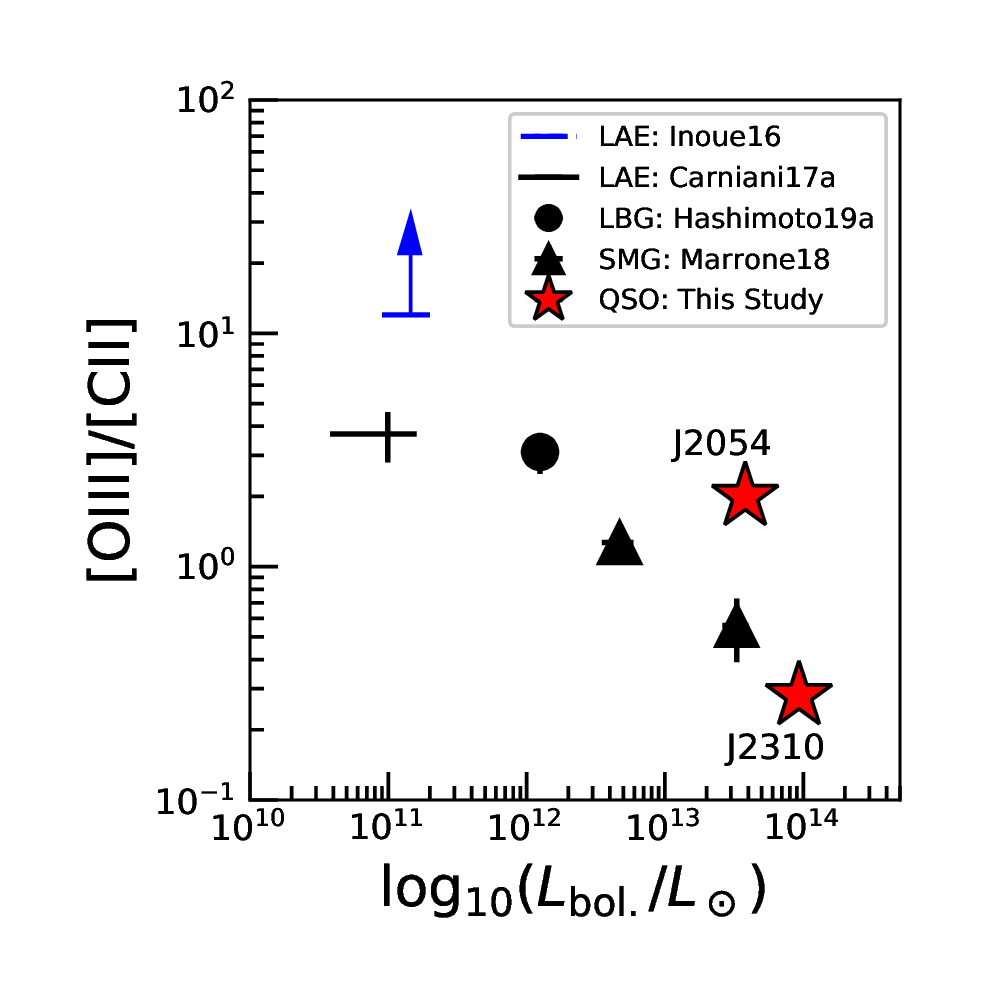}

\caption
{
{\bf (Left panel)} The $L_{\rm [OIII]}$-to-\ltir\ ratio plotted against \ltir, where the luminosities are corrected for magnification, if any. 
Orange hexagons show the two quasars. 
The other eight filled symbols represent  $z\approx7-9$ objects compiled by \cite{tamura2019}: 
SPT0311$-$58 E/W at $z=6.90$ (triangles, \citealt{marrone2018}); 
BDF-3299 at $z=7.11$ (pentagon with two arrows, \citealt{carniani2017a}), 
B14-65666 at $z=7.15$ (filled diamond, \citealt{hashimoto2019a}), 
SXDF-NB1006-2 at $z=7.21$ (circle with two arrows, \citealt{inoue2016}), 
MACS0416$\_$Y1 at $z=8.31$ (five-pointed star, \citealt{tamura2019}), 
A2744$\_$YD4 at $z=8.38$ (square, \citealt{laporte2017}), 
and MACS1149-JD1 at $z=9.11$ (triangle with two arrows, \citealt{hashimoto2018}).  
The open symbols show lower-$z$ galaxies including 
the Herschel DGS (open circles, \citealt{cormier2015}) 
and SHINING samples (thin open squares, \citealt{herrera-camus2018a}), 
the median of local spirals (thick open square, \citealt{brauher2008}), 
and $z\approx2-4$ dusty star-forming galaxies with spectroscopic redshifts (open triangles, \citealt{zhang2018}). 
For the $z>6$ objects except for the two quasars and SPT0311$-$58 E/W, we have assumed \ltir\ $=$ 50 K and $\beta_{\rm d} = 1.6$ for consistency. 
The blue-to-green color code shown for MACS0416$\_$Y1, SXDF-NB1006-2, MACS1149-JD1, B14-65666 and local dwarfs indicates the best-fitting oxygen abundances.
{\bf (Right panel)} The \oiii-to-\cii\ luminosity ratio at $z>6$. 
The sample includes the two quasars (two red five-pointed stars); 
SPT0311$-$58 E/W (triangles); 
B14-65666 (circle); 
BDF-3299 ('+' symbol); 
and SXDF-NB1006-2 ('-' symbol with an upward arrow). 
Note that the definition of $L_{\rm bol.}$ is different for quasars and star-forming galaxies: $L_{\rm bol.}$ of quasars indicate the quasar/AGN bolometric luminosity, which are taken from the literature. On the other hand, $L_{\rm bol.}$ of other galaxies without AGN activity indicate the galaxy's bolometric luminosity.} These are obtained in \cite{hashimoto2019a} as the summation of the UV luminosity and \ltir, where we assume \ltir\ $=$ 50 K and $\beta_{\rm d} = 1.6$ except for SPT0311$-$58 E/W. For the two LAEs with \ltir\ upper limits, the lower limit corresponds to the UV luminosity, while the upper limit denotes the summation of the UV luminosity and the $3\sigma$ \ltir\ upper limits. 
\label{fig3}
\end{figure*}

It is widely known that the FIR line luminosity-to-\ltir\ ratio anti-correlates with \ltir, the so-called FIR line deficit, particularly at the high \ltir\ regime (\ltir\ $\gtrsim10^{11}$ \lsun) (e.g., \citealt{malhotra1997, stacey2010, gracia-carpio2011, diaz.santos2013, diaz.santos2017, herrera-camus2018a, herrera-camus2018b}).
The line deficit has been first identified in \cii. A number of hypotheses are proposed to explain the \cii\ line deficit in extreme objects such as (ultra-)luminous infrared galaxies, (U)LIRGs, and luminous quasars, although a consensus is yet to be reached (e.g.,  \citealt{kaufman1999, malhotra2001, luhman2003, abel2009, gracia-carpio2011, langer.pineda2015, munoz.oh2016, diaz.santos2017, lagache2018, herrera-camus2018a, rybak2019}). 

Focusing on \oiii, based on a compiled sample of local dwarf and spiral galaxies with high dynamic ranges in metallicity and \ltir, \cite{cormier2015} have shown that the $L_{\rm [OIII]}$-to-\ltir\ ratio anti-correlates with \ltir\ (see also e.g., \citealt{malhotra1997, gracia-carpio2011, diaz.santos2017, herrera-camus2018a, herrera-camus2018b}). The local galaxies have the $L_{\rm [OIII]}$-to-\ltir\ ratio ranging from $\approx 10^{-5}$ to $\approx 10^{-2}$. 
Recently, \cite{tamura2019} have investigated the relation at higher-$z$ based on a compiled sample of  $z\approx7-9$ galaxies, showing that at least high-$z$ galaxies with dust continuum detections follow a similar relation as in the local Universe. Our two quasars are useful to further investigate the trend at the reionization epoch because of their high \ltir\ values. The luminosity ratios are log ($L_{\rm [OIII]}$/\ltir) $=-3.3\pm0.1$ and $-4.0\pm0.1$ in \namea\ and \nameb, respectively. 

In the left panel of Figure \ref{fig3}, we plot the two quasars along with eight objects at $z > 7$ (see caption for the details) and lower-$z$ objects. The latter includes various populations of local galaxies taken from the Herschel DGS (\citealt{cormier2015}) and SHINING  (\citealt{herrera-camus2018a}) samples, local spirals (\citealt{brauher2008}), and lensed sub-millimeter galaxies (SMGs) at $z\approx1-4$ (\citealt{zhang2018}). We confirm a trend that high-$z$ objects follow a similar relation as in the local Universe. 
Following explanations for the \cii\ line deficit, we propose two possible explanations for the \oiii\ line deficit.Firstly, the collisional de-excitation of \oiii\ may significantly reduce $L_{\rm [OIII]}$ (and hence the luminosity ratio) in the high electron density environment ($n_{\rm e} > n_{\rm crit.} \approx 500$ cm$^{-3}$). Secondly, the strong AGN radiation can contribute to \ltir\ (e.g., \citealt{symeonidis2016}) while ionize oxygen higher than O$^{2+}$ (e.g., \citealt{spinoglio2015}). Detailed modeling is needed to conclude the origin of the \oiii-line deficit, which we leave for future studies. 

In the left panel of Figure \ref{fig3}, we find that the relation between $L_{\rm [OIII]}$-to-\ltir\ and \ltir\ at high-$z$ is shifted toward higher \ltir\ values (or higher $L_{\rm [OIII]}$-to-\ltir\ values). Such a shift toward higher \ltir\ values is also found in the relation between $L_{\rm [CII]}$-to-\ltir\ and \ltir\ (e.g., \citealt{stacey2010}). In the case of \cii, to understand the origin of this shift in high-$z$ galaxies, \cite{narayanan.krumholz2017} have coupled analytical models for the structure of giant molecular clouds in galaxies with chemical equilibrium networks and radiative transfer models. The authors have proposed that the shift could arise if high-$z$ galaxies have larger gas masses at a given SFR (i.e, \ltir). In this case, the gas surface density of an individual molecular cloud (i.e. the star formation efficiency, SFE $= SFR/M_{\rm mol}$, defined as the SFR per molecular gas mass) becomes lower, which in turn reduces the ability for self-shielding of molecular clouds against ionizing photons, leading to a brighter \cii\ luminosity. Although it is not clear whether the same discussion is applicable for \oiii, there is growing evidence that the molecular gas properties are the key to regulate the general trend in the FIR line deficit; \cite{herrera-camus2018a} shows that the discrepancy between low- and high-$z$ objects becomes less prominent if one plots the \lcii-to-\ltir\ ratio against the IR surface brightness ($\Sigma_{\rm IR}$) , the latter of which is related to the gas surface density or the SFE. Such an analysis for \oiii\ would be interesting in future with larger samples in local and high-$z$ Universe. 



\subsubsection{\oiii-to-\cii\ Luminosity Ratio}
\label{subsubsec:lumi_ratio2}

We next turn our attention to the \oiii-to-\cii\ luminosity ratio. Based on a compiled sample of five galaxies at $z \gtrsim 7$ with \oiii\ and \cii\ observations (\citealt{inoue2016, carniani2017a, marrone2018}), \cite{hashimoto2019a} have demonstrated a trend that the \oiii-to-\cii\ line luminosity ratio becomes small if a galaxy has a large galaxy bolometric luminosity. Their sample includes two Ly$\alpha$ emitters (LAEs), one Lyman break galaxy (LBG), and two SMGs. \namea\ and \nameb\ offers us an invaluable opportunity to investigate the line luminosity ratio in quasars at $z>6$ for the first time. 

In \namea\ (\nameb), combining our \oiii\ luminosity and the \cii\ luminosity of $3.3\pm0.5\times10^{9}$ \lsun\ ($8.7\pm1.4\times10^{9}$ \lsun) in \cite{wang2013}, we obtain the line luminosity ratio of  $2.1\pm0.4$ ($0.3\pm0.1$). In the right panel of Figure \ref{fig3}, red star symbols show the line luminosity ratio of the two quasars plotted against the {\it quasar/AGN} bolometric luminosity. Just for comparisons, black and blue symbols show the line luminosity ratios in star forming galaxies without AGN activity at $z>6$ plotted against the {\it galaxy} bolometric luminosity. Notably, \nameb\ has the lowest \oiii-to-\cii\ ratio so far reported among objects at $z>6$.



\section{Discussion and Summary}
\label{sec:discussion}

To interpret \oiii\ in quasars, we need to separate the \oiii\ contribution from star formation and AGN activity. For the latter, \oiii\ can arise from the Narrow Line Region (NLR) of AGNs because the NLR has a relatively small electron density of $100-300$ cm$^{-3}$ (e.g., \citealt{bennert2006, kakkad2018}), which is smaller than the critical density of \oiii\ 88 \micron\ ($\approx 500$ cm$^{-3}$). Indeed, the size of the \oiii\ emitting region of the two quasars ($\approx2-3$ kpc in FWHM) are reasonable for the size of stellar disks or extended NLRs. 
Ideally, one can separate the contribution based on spatially resolved diagnostics such as the BPT diagram (\citealt{baldwin1981}) as performed in local Seyfert and LINER galaxies (\citealt{kakkad2018}).  However, it is difficult to separate the contribution based on the \oiii-to-\cii\ line ratio alone because it is insensitive to the presence of AGNs;  \cite{herrera-camus2018a}  demonstrate that both star formation and AGN activity can reproduce the \oiii-to-\cii\ luminosity ratio of $\approx0.1 - 2.0$ (see their Figure 11). 

We thus focus on the fact that \oiii\ and \cii\ have consistent redshifts, emitting region sizes, and line FWHMs within uncertainties (\S \ref{subsec:oiii}). These results would imply that \oiii\  arises from star-forming regions as traced by \cii\ (see similar argument by \citealt{walter2018}), although we cannot rule out the possibility that both \oiii\ and \cii\ are partly affected by the NLRs. 

We next examine if \oiii\ luminosity-based SFRs (SFR$_{\rm OIII}$) are comparable to SFR$_{\rm IR}$ values (Table \ref{tab1}). It is expected that the two SFR values are consistent with each other if \oiii\ is mainly powered by star-forming acticity. For the conversion of the \oiii\ luminosity to the SFR, we use the empirical relation in the local Universe which assumes the Kroupa IMF in the range of $0.1-100$ \msun\ (\citealt{de.looze2014}). The authors present different empirical relations for e.g., metal-poor dwarf galaxies, starburst, the composite of star formation and AGNs, ULIRGs, and the entire sample. 
Because the two quasars presented in this study have \ltir\ $\approx10^{13}$ \lsun, the empirical relation for ULIRGs would be suitable among the relations in \cite{de.looze2014}. Although the typical uncertainty of the relation is a factor of 2.5 (see Table 3 in \citealt{de.looze2014}), it is unclear whether the local relation can be applied for luminous quasars at $z=6$ presented in this study. Therefore, the actual uncertainty would be larger than a factor of 2.5. On the other hand, as discussed in \S \ref{subsec:discussion1}, the SFR values estimated from \ltir\ are also highly uncertain because it is possible that AGN significantly contaminates the FIR dust continuum emission (e.g., \citealt{symeonidis2016}). With these in mind, we find that  SFR$_{\rm OIII}$ in \namea\ is about five times larger than SFR$_{\rm IR}$, whereas the two SFR values in \nameb\ are consistent within a factor of two. Given the large systematic uncertainties in the two SFR values, there is no strong evidence that supports the AGN contamination to \oiii\ in the two objects in this study\footnote{\cite{walter2018} have also compared the two SFR values to examine the origin of \oiii. Using the empirical relation for high-$z$ ($z>0.5$) galaxies, the authors argue that the two SFRs are in good agreement. However, the \oiii\ empirical relation for high-$z$ galaxies is constructed only with three objects (see Table 3 in \citealt{de.looze2014}). Thus, their comparison is also subject to the uncertainties as described in this study.}.

A possible way to disentangle the \oiii\ contributions from star formation and AGN activity is to investigate a spatially resolved map of \oi\ 63 \micron-to-\cii\ line ratio defining the AGN-dominated region. This is because \oi\ is significantly enhanced in the presence of AGNs due to the fact that \oi\ becomes a more efficient coolant than \cii\ in dense and warmer gas (\citealt{herrera-camus2018a}). Combining this map with high-angular resolution \oiii\ data, future works can infer the \oiii\ flux fraction of each component. 

Although the origin of \oiii\ emission is not clear given the current data, we try to interpret the \oiii-to-\cii\ line luminosity ratio of the two quasars. In the local Universe, based on a compiled sample of star-forming galaxies and AGN-dominated galaxies, \cite{herrera-camus2018a} have statistically demonstrated that \oiii\ becomes stronger than \cii\ if  galaxies have higher dust temperature (see also \citealt{diaz.santos2017}). The two quasars seem to be consistent with the trend in the sense that \namea\ (\nameb) has high (low) dust temperature, $T_{\rm d} = 50\pm2$ K ($37\pm1$ K). An interpretation of the result is that \namea\ has a harder UV stellar + AGN radiation field than \nameb\footnote{In the case of \namea, AGN activity may significantly contribute to the radiation field because the object has very compact dust emitting region.}. 
Assuming the same dust covering fraction and the dust grain size distribution, a harder UV radiation field leads to higher \td. The harder UV radiation field also naturally enhances \oiii\ (ionization potential $\approx$ 35 eV) against \cii\ (ionization potential $\approx$ 11 eV) if we assume a constant C/O abundance ratio. This hypothesis can be tested with the line luminosity ratio of \nii\ 205 \micron\ against \oiii, which is a good tracer of the UV radiation hardness (\citealt{ferkinhoff2010}). Alternatively, the weak \oiii\ in \nameb\ may be due to its high electron density that causes collisional de-excitation. This can be investigated with the line ratio of \oiii\ 88 \micron-to-\oiii\ 52 \micron, which is sensitive to the electron density because of their different critical densities (\citealt{pereira-sataella2017}). 
Our results highlight the potential use of \oiii\ (and the underlying continuum) as a useful tracer of the ISM in the quasar host galaxies.

\section*{Acknowledgments}

We thank an anonymous referee for valuable comments that have greatly improved the paper. 
We are grateful to Toru Nagao, Nobunari Kashikawa, Yoshiki Matsuoka, Kohei Ichikawa, and Takuma Izumi for discussion. 
This paper makes use of the following ALMA data: 
ADS/JAO.ALMA\#2017.1.01195.S. ALMA is a partnership of ESO (representing its member states), 
NSF (USA) and NINS (Japan), together with NRC (Canada), NSC and ASIAA (Taiwan), and KASI (Republic of Korea), 
in cooperation with the Republic of Chile. 
The Joint ALMA Observatory is operated by ESO, AUI/NRAO and NAOJ.

Funding for SDSS-III has been provided by the Alfred P. Sloan Foundation, the Participating Institutions, the National Science Foundation, and the U.S. Department of Energy Office of Science. The SDSS-III web site is http://www.sdss3.org/.
SDSS-III is managed by the Astrophysical Research Consortium for the Participating Institutions of the SDSS-III Collaboration including the University of Arizona, the Brazilian Participation Group, Brookhaven National Laboratory, Carnegie Mellon University, University of Florida, the French Participation Group, the German Participation Group, Harvard University, the Instituto de Astrofisica de Canarias, the Michigan State/Notre Dame/JINA Participation Group, Johns Hopkins University, Lawrence Berkeley National Laboratory, Max Planck Institute for Astrophysics, Max Planck Institute for Extraterrestrial Physics, New Mexico State University, New York University, Ohio State University, Pennsylvania State University, University of Portsmouth, Princeton University, the Spanish Participation Group, University of Tokyo, University of Utah, Vanderbilt University, University of Virginia, University of Washington, and Yale University.

This work has made use of data from the European Space Agency (ESA) mission
{\it Gaia} (\url{https://www.cosmos.esa.int/gaia}), processed by the {\it Gaia}
Data Processing and Analysis Consortium (DPAC,
\url{https://www.cosmos.esa.int/web/gaia/dpac/consortium}). Funding for the DPAC
has been provided by national institutions, in particular the institutions
participating in the {\it Gaia} Multilateral Agreement.

T.H. and A.K.I. appreciate support from NAOJ ALMA Scientific Research Grant Number 2016-01A. 
We are also grateful to KAKENHI grants 26287034 and 17H01114 (K.M. and A.K.I.), 17H06130 (Y.T.). 
This work was partly supported by the Grant-inAid for Scientific Research 19J01620 (T.H.).

We thank Rodrigo Herrera-Camus for kindly providing us with their SHINING sample data used in the left panel of Figure \ref{fig3}. 


\begin{thebibliography}{79}
\expandafter\ifx\csname natexlab\endcsname\relax\def\natexlab#1{#1}\fi

\bibitem[{{Abel} {et~al.}(2009){Abel}, {Dudley}, {Fischer}, {Satyapal}, \& {van
  Hoof}}]{abel2009}
{Abel}, N.~P., {Dudley}, C., {Fischer}, J., {Satyapal}, S., \& {van Hoof},
  P.~A.~M. 2009, \apj, 701, 1147

\bibitem[{{Ba{\~n}ados} {et~al.}(2015){Ba{\~n}ados}, {Decarli}, {Walter},
  {Venemans}, {Farina}, \& {Fan}}]{banados2015}
{Ba{\~n}ados}, E., {Decarli}, R., {Walter}, F., {Venemans}, B.~P., {Farina},
  E.~P., \& {Fan}, X. 2015, \apjl, 805, L8

\bibitem[{{Ba{\~n}ados} {et~al.}(2018){Ba{\~n}ados}, {Venemans},
  {Mazzucchelli}, {Farina}, {Walter}, {Wang}, {Decarli}, {Stern}, {Fan},
  {Davies}, {Hennawi}, {Simcoe}, {Turner}, {Rix}, {Yang}, {Kelson}, {Rudie}, \&
  {Winters}}]{banados2018}
{Ba{\~n}ados}, E., {et~al.} 2018, \nat, 553, 473

\bibitem[{{Baldwin} {et~al.}(1981){Baldwin}, {Phillips}, \&
  {Terlevich}}]{baldwin1981}
{Baldwin}, J.~A., {Phillips}, M.~M., \& {Terlevich}, R. 1981, \pasp, 93, 5

\bibitem[{{Beelen} {et~al.}(2006){Beelen}, {Cox}, {Benford}, {Dowell},
  {Kov{\'a}cs}, {Bertoldi}, {Omont}, \& {Carilli}}]{beelen2006}
{Beelen}, A., {Cox}, P., {Benford}, D.~J., {Dowell}, C.~D., {Kov{\'a}cs}, A.,
  {Bertoldi}, F., {Omont}, A., \& {Carilli}, C.~L. 2006, \apj, 642, 694

\bibitem[{{Bennert} {et~al.}(2006){Bennert}, {Jungwiert}, {Komossa}, {Haas}, \&
  {Chini}}]{bennert2006}
{Bennert}, N., {Jungwiert}, B., {Komossa}, S., {Haas}, M., \& {Chini}, R. 2006,
  \aap, 456, 953

\bibitem[{{Brauher} {et~al.}(2008){Brauher}, {Dale}, \& {Helou}}]{brauher2008}
{Brauher}, J.~R., {Dale}, D.~A., \& {Helou}, G. 2008, \apjs, 178, 280

\bibitem[{{Carniani} {et~al.}(2017){Carniani}, {Maiolino}, {Pallottini},
  {Vallini}, {Pentericci}, {Ferrara}, {Castellano}, {Vanzella}, {Grazian},
  {Gallerani}, {Santini}, {Wagg}, \& {Fontana}}]{carniani2017a}
{Carniani}, S., {et~al.} 2017, \aap, 605, A42

\bibitem[{{Cormier} {et~al.}(2015){Cormier}, {Madden}, {Lebouteiller}, {Abel},
  {Hony}, {Galliano}, {R{\'e}my-Ruyer}, {Bigiel}, {Baes}, {Boselli},
  {Chevance}, {Cooray}, {De Looze}, {Doublier}, {Galametz}, {Hughes},
  {Karczewski}, {Lee}, {Lu}, \& {Spinoglio}}]{cormier2015}
{Cormier}, D., {et~al.} 2015, \aap, 578, A53

\bibitem[{{da Cunha} {et~al.}(2013){da Cunha}, {Groves}, {Walter}, {Decarli},
  {Weiss}, {Bertoldi}, {Carilli}, {Daddi}, {Elbaz}, {Ivison}, {Maiolino},
  {Riechers}, {Rix}, {Sargent}, \& {Smail}}]{da_cunha2013}
{da Cunha}, E., {et~al.} 2013, \apj, 766, 13

\bibitem[{{De Looze} {et~al.}(2014){De Looze}, {Cormier}, {Lebouteiller},
  {Madden}, {Baes}, {Bendo}, {Boquien}, {Boselli}, {Clements}, {Cortese},
  {Cooray}, {Galametz}, {Galliano}, {Graci{\'a}-Carpio}, {Isaak}, {Karczewski},
  {Parkin}, {Pellegrini}, {R{\'e}my-Ruyer}, {Spinoglio}, {Smith}, \&
  {Sturm}}]{de.looze2014}
{De Looze}, I., {et~al.} 2014, \aap, 568, A62

\bibitem[{{De Rosa} {et~al.}(2014){De Rosa}, {Venemans}, {Decarli}, {Gennaro},
  {Simcoe}, {Dietrich}, {Peterson}, {Walter}, {Frank}, {McMahon}, {Hewett},
  {Mortlock}, \& {Simpson}}]{de.rosa2014}
{De Rosa}, G., {et~al.} 2014, \apj, 790, 145

\bibitem[{{Decarli} {et~al.}(2018){Decarli}, {Walter}, {Venemans},
  {Ba{\~n}ados}, {Bertoldi}, {Carilli}, {Fan}, {Farina}, {Mazzucchelli},
  {Riechers}, {Rix}, {Strauss}, {Wang}, \& {Yang}}]{decarli2018}
{Decarli}, R., {et~al.} 2018, \apj, 854, 97

\bibitem[{{D{\'{\i}}az-Santos} {et~al.}(2013){D{\'{\i}}az-Santos}, {Armus},
  {Charmandaris}, {Stierwalt}, {Murphy}, {Haan}, {Inami}, {Malhotra},
  {Meijerink}, {Stacey}, {Petric}, {Evans}, {Veilleux}, {van der Werf}, {Lord},
  {Lu}, {Howell}, {Appleton}, {Mazzarella}, {Surace}, {Xu}, {Schulz},
  {Sanders}, {Bridge}, {Chan}, {Frayer}, {Iwasawa}, {Melbourne}, \&
  {Sturm}}]{diaz.santos2013}
{D{\'{\i}}az-Santos}, T., {et~al.} 2013, \apj, 774, 68

\bibitem[{{D{\'{\i}}az-Santos} {et~al.}(2017){D{\'{\i}}az-Santos}, {Armus},
  {Charmandaris}, {Lu}, {Stierwalt}, {Stacey}, {Malhotra}, {van der Werf},
  {Howell}, {Privon}, {Mazzarella}, {Goldsmith}, {Murphy}, {Barcos-Mu{\~n}oz},
  {Linden}, {Inami}, {Larson}, {Evans}, {Appleton}, {Iwasawa}, {Lord},
  {Sanders}, \& {Surace}}]{diaz.santos2017}
{D{\'{\i}}az-Santos}, T., {et~al.}  2017, \apj, 846, 32

\bibitem[{{Eisenstein} {et~al.}(2011){Eisenstein}, {Weinberg}, {Agol},
  {Aihara}, {Allende Prieto}, {Anderson}, {Arns}, {Aubourg}, {Bailey},
  {Balbinot}, \& et~al.}]{eisenstein2011}
{Eisenstein}, D.~J., {et~al.} 2011, \aj, 142, 72

\bibitem[{{Fan} {et~al.}(2003){Fan}, {Strauss}, {Schneider}, {Becker}, {White},
  {Haiman}, {Gregg}, {Pentericci}, {Grebel}, {Narayanan}, {Loh}, {Richards},
  {Gunn}, {Lupton}, {Knapp}, {Ivezi{\'c}}, {Brandt}, {Collinge}, {Hao},
  {Harbeck}, {Prada}, {Schaye}, {Strateva}, {Zakamska}, {Anderson},
  {Brinkmann}, {Bahcall}, {Lamb}, {Okamura}, {Szalay}, \& {York}}]{fan2003}
{Fan}, X., {et~al.} 2003, \aj, 125, 1649

\bibitem[{{Ferkinhoff} {et~al.}(2010){Ferkinhoff}, {Hailey-Dunsheath},
  {Nikola}, {Parshley}, {Stacey}, {Benford}, \& {Staguhn}}]{ferkinhoff2010}
{Ferkinhoff}, C., {Hailey-Dunsheath}, S., {Nikola}, T., {Parshley}, S.~C.,
  {Stacey}, G.~J., {Benford}, D.~J., \& {Staguhn}, J.~G. 2010, \apjl, 714, L147

\bibitem[{{Feruglio} {et~al.}(2018){Feruglio}, {Fiore}, {Carniani}, {Maiolino},
  {D'Odorico}, {Luminari}, {Barai}, {Bischetti}, {Bongiorno}, {Cristiani},
  {Ferrara}, {Gallerani}, {Marconi}, {Pallottini}, {Piconcelli}, \&
  {Zappacosta}}]{feruglio2018}
{Feruglio}, C., {et~al.} 2018, \aa, 619A, 39

\bibitem[{{Gaia Collaboration} {et~al.}(2016){Gaia Collaboration}, {Brown},
  {Vallenari}, {Prusti}, {de Bruijne}, {Mignard}, {Drimmel}, {Babusiaux},
  {Bailer-Jones}, {Bastian}, \& et~al.}]{gaia2016}
{Gaia Collaboration} {et~al.} 2016, \aap, 595, A2

\bibitem[{{Gaia Collaboration} {et~al.}(2018){Gaia Collaboration}, {Brown},
  {Vallenari}, {Prusti}, {de Bruijne}, {Babusiaux}, {Bailer-Jones}, {Biermann},
  {Evans}, {Eyer}, \& et~al.}]{gaia2018}
{Gaia Collaboration} {et~al.} 2018, \aap, 616, A1

\bibitem[{{Graci{\'a}-Carpio} {et~al.}(2011){Graci{\'a}-Carpio}, {Sturm},
  {Hailey-Dunsheath}, {Fischer}, {Contursi}, {Poglitsch}, {Genzel},
  {Gonz{\'a}lez-Alfonso}, {Sternberg}, {Verma}, {Christopher}, {Davies},
  {Feuchtgruber}, {de Jong}, {Lutz}, \& {Tacconi}}]{gracia-carpio2011}
{Graci{\'a}-Carpio}, J., {et~al.} 2011, \apjl, 728, L7

\bibitem[{{H{\"a}ring} \& {Rix}(2004)}]{haering2004}
{H{\"a}ring}, N., \& {Rix}, H.-W. 2004, \apjl, 604, L89

\bibitem[{{Hashimoto} {et~al.}(2019){Hashimoto}, {Inoue},
  {Mawatari}, {Tamura}, {Matsuo}, {Furusawa}, {Harikane}, {Shibuya}, {Knudsen},
  {Kohno}, {Ono}, {Zackrisson}, {Okamoto}, {Kashikawa}, {Oesch}, {Ouchi},
  {Ota}, {Shimizu}, {Taniguchi}, {Umehata}, \& {Watson}}]{hashimoto2019a}
{Hashimoto}, T., {et~al.} 2019, PASJ in press (arXiv:1806.00486)

\bibitem[{{Hashimoto} {et~al.}(2018){Hashimoto}, {Laporte},
  {Mawatari}, {Ellis}, {Inoue}, {Zackrisson}, {Roberts-Borsani}, {Zheng},
  {Tamura}, {Bauer}, {Fletcher}, {Harikane}, {Hatsukade}, {Hayatsu}, {Matsuda},
  {Matsuo}, {Okamoto}, {Ouchi}, {Pell{\'o}}, {Rydberg}, {Shimizu}, {Taniguchi},
  {Umehata}, \& {Yoshida}}]{hashimoto2018}
{Hashimoto}, T., {et~al.} 2018, \nat, 557, 392

\bibitem[{{Herrera-Camus} {et~al.}(2018{\natexlab{a}}){Herrera-Camus}, {Sturm},
  {Graci{\'a}-Carpio}, {Lutz}, {Contursi}, {Veilleux}, {Fischer},
  {Gonz{\'a}lez-Alfonso}, {Poglitsch}, {Tacconi}, {Genzel}, {Maiolino},
  {Sternberg}, {Davies}, \& {Verma}}]{herrera-camus2018a}
{Herrera-Camus}, R., {et~al.} 2018{\natexlab{a}}, \apj, 861, 94

\bibitem[{{Herrera-Camus} {et~al.}(2018{\natexlab{b}}){Herrera-Camus}, {Sturm},
  {Graci{\'a}-Carpio}, {Lutz}, {Contursi}, {Veilleux}, {Fischer},
  {Gonz{\'a}lez-Alfonso}, {Poglitsch}, {Tacconi}, {Genzel}, {Maiolino},
  {Sternberg}, {Davies}, \& {Verma}}]{herrera-camus2018b}
{Herrera-Camus}, R., {et~al.} 2018{\natexlab{b}}, \apj, 861, 95

\bibitem[{{Ichikawa} {et~al.}(2019){Ichikawa}, {Ricci}, {Ueda}, {Bauer},
  {Kawamuro}, {Koss}, {Oh}, {Rosario}, {Shimizu}, {Stalevski}, {Fuller},
  {Packham}, \& {Trakhtenbrot}}]{ichikawa2019}
{Ichikawa}, K., {et~al.} 2019, \apj, 870, 31

\bibitem[{{Inoue} {et~al.}(2016){Inoue}, {Tamura}, {Matsuo}, {Mawatari},
  {Shimizu}, {Shibuya}, {Ota}, {Yoshida}, {Zackrisson}, {Kashikawa}, {Kohno},
  {Umehata}, {Hatsukade}, {Iye}, {Matsuda}, {Okamoto}, \&
  {Yamaguchi}}]{inoue2016}
{Inoue}, A.~K., {et~al.} 2016, Science, 352, 1559

\bibitem[{{Izumi} {et~al.}(2018){Izumi}, {Onoue}, {Shirakata}, {Nagao},
  {Kohno}, {Matsuoka}, {Imanishi}, {Strauss}, {Kashikawa}, {Schulze},
  {Silverman}, {Fujimoto}, {Harikane}, {Toba}, {Umehata}, {Nakanishi},
  {Greene}, {Tamura}, {Taniguchi}, {Yamaguchi}, {Goto}, {Hashimoto},
  {Ikarashi}, {Iono}, {Iwasawa}, {Lee}, {Makiya}, {Minezaki}, \&
  {Tang}}]{izumi2018}
{Izumi}, T., {et~al.} 2018, \pasj, 70, 36

\bibitem[{{Jiang} {et~al.}(2008){Jiang}, {Fan}, {Annis}, {Becker}, {White},
  {Chiu}, {Lin}, {Lupton}, {Richards}, {Strauss}, {Jester}, \&
  {Schneider}}]{jiang2008}
{Jiang}, L., {et~al.} 2008, \aj, 135, 1057

\bibitem[{{Jiang} {et~al.}(2016){Jiang}, {McGreer}, {Fan}, {Strauss},
  {Ba{\~n}ados}, {Becker}, {Bian}, {Farnsworth}, {Shen}, {Wang}, {Wang},
  {Wang}, {White}, {Wu}, {Wu}, {Yang}, \& {Yang}}]{jiang2016}
{Jiang}, L., {et~al.} 2016, \apj, 833, 222

\bibitem[{{Kakkad} {et~al.}(2018){Kakkad}, {Groves}, {Dopita}, {Thomas},
  {Davies}, {Mainieri}, {Kharb}, {Scharw{\"a}chter}, {Hampton}, \&
  {Ho}}]{kakkad2018}
{Kakkad}, D., {et~al.} 2018, \aap, 618, A6

\bibitem[{{Kaufman} {et~al.}(1999){Kaufman}, {Wolfire}, {Hollenbach}, \&
  {Luhman}}]{kaufman1999}
{Kaufman}, M.~J., {Wolfire}, M.~G., {Hollenbach}, D.~J., \& {Luhman}, M.~L.
  1999, \apj, 527, 795

\bibitem[{{Kennicutt} \& {Evans}(2012)}]{kennicutt.evans2012}
{Kennicutt}, R.~C., \& {Evans}, N.~J. 2012, \araa, 50, 531

\bibitem[{{Komatsu} {et~al.}(2011){Komatsu}, {Smith}, {Dunkley}, {Bennett},
  {Gold}, {Hinshaw}, {Jarosik}, {Larson}, {Nolta}, {Page}, {Spergel},
  {Halpern}, {Hill}, {Kogut}, {Limon}, {Meyer}, {Odegard}, {Tucker}, {Weiland},
  {Wollack}, \& {Wright}}]{komatsu2011}
{Komatsu}, E., {et~al.} 2011, \apjs, 192, 18

\bibitem[{{Kormendy} \& {Ho}(2013)}]{kormendy.ho2013}
{Kormendy}, J., \& {Ho}, L.~C. 2013, \araa, 51, 511

\bibitem[{{Kroupa}(2001)}]{kroupa2001}
{Kroupa}, P. 2001, \mnras, 322, 231

\bibitem[{{Lagache} {et~al.}(2018){Lagache}, {Cousin}, \&
  {Chatzikos}}]{lagache2018}
{Lagache}, G., {Cousin}, M., \& {Chatzikos}, M. 2018, \aap, 609, A130

\bibitem[{{Langer} \& {Pineda}(2015)}]{langer.pineda2015}
{Langer}, W.~D., \& {Pineda}, J.~L. 2015, \aap, 580, A5

\bibitem[{{Laporte} {et~al.}(2017){Laporte}, {Ellis}, {Boone}, {Bauer},
  {Qu{\'e}nard}, {Roberts-Borsani}, {Pell{\'o}}, {P{\'e}rez-Fournon}, \&
  {Streblyanska}}]{laporte2017}
{Laporte}, N., {et~al.} 2017, \apjl, 837, L21

\bibitem[{{Leipski} {et~al.}(2013){Leipski}, {Meisenheimer}, {Walter}, {Besel},
  {Dannerbauer}, {Fan}, {Haas}, {Klaas}, {Krause}, \& {Rix}}]{leipski2013}
{Leipski}, C., {et~al.} 2013, \apj, 772, 103

\bibitem[{{Luhman} {et~al.}(2003){Luhman}, {Satyapal}, {Fischer}, {Wolfire},
  {Sturm}, {Dudley}, {Lutz}, \& {Genzel}}]{luhman2003}
{Luhman}, M.~L., {Satyapal}, S., {Fischer}, J., {Wolfire}, M.~G., {Sturm}, E.,
  {Dudley}, C.~C., {Lutz}, D., \& {Genzel}, R. 2003, \apj, 594, 758

\bibitem[{{Maiolino} {et~al.}(2005){Maiolino}, {Cox}, {Caselli}, {Beelen},
  {Bertoldi}, {Carilli}, {Kaufman}, {Menten}, {Nagao}, {Omont}, {Wei{\ss}},
  {Walmsley}, \& {Walter}}]{maiolino2005}
{Maiolino}, R., {et~al.} 2005, \aap, 440, L51

\bibitem[{{Maiolino} {et~al.}(2012){Maiolino}, {Gallerani}, {Neri}, {Cicone},
  {Ferrara}, {Genzel}, {Lutz}, {Sturm}, {Tacconi}, {Walter}, {Feruglio},
  {Fiore}, \& {Piconcelli}}]{maiolino2012}
{Maiolino}, R., {et~al.} 2012, \mnras, 425, L66

\bibitem[{{Malhotra} {et~al.}(1997){Malhotra}, {Helou}, {Stacey}, {Hollenbach},
  {Lord}, {Beichman}, {Dinerstein}, {Hunter}, {Lo}, {Lu}, {Rubin},
  {Silbermann}, {Thronson}, \& {Werner}}]{malhotra1997}
{Malhotra}, S., {et~al.} 1997, \apjl, 491, L27

\bibitem[{{Malhotra} {et~al.}(2001){Malhotra}, {Kaufman}, {Hollenbach},
  {Helou}, {Rubin}, {Brauher}, {Dale}, {Lu}, {Lord}, {Stacey}, {Contursi},
  {Hunter}, \& {Dinerstein}}]{malhotra2001}
{Malhotra}, S., {et~al.} 2001, \apj, 561, 766

\bibitem[{{Marrone} {et~al.}(2018){Marrone}, {Spilker}, {Hayward}, {Vieira},
  {Aravena}, {Ashby}, {Bayliss}, {B{\'e}thermin}, {Brodwin}, {Bothwell},
  {Carlstrom}, {Chapman}, {Chen}, {Crawford}, {Cunningham}, {De Breuck},
  {Fassnacht}, {Gonzalez}, {Greve}, {Hezaveh}, {Lacaille}, {Litke}, {Lower},
  {Ma}, {Malkan}, {Miller}, {Morningstar}, {Murphy}, {Narayanan}, {Phadke},
  {Rotermund}, {Sreevani}, {Stalder}, {Stark}, {Strandet}, {Tang}, \&
  {Wei{\ss}}}]{marrone2018}
{Marrone}, D.~P., {et~al.} 2018, \nat, 553, 51

\bibitem[{{Matsuoka} {et~al.}(2018){Matsuoka}, {Iwasawa}, {Onoue}, {Kashikawa},
  {Strauss}, {Lee}, {Imanishi}, {Nagao}, {Akiyama}, {Asami}, {Bosch},
  {Furusawa}, {Goto}, {Gunn}, {Harikane}, {Ikeda}, {Izumi}, {Kawaguchi},
  {Kato}, {Kikuta}, {Kohno}, {Komiyama}, {Lupton}, {Minezaki}, {Miyazaki},
  {Morokuma}, {Murayama}, {Niida}, {Nishizawa}, {Oguri}, {Ono}, {Ouchi},
  {Price}, {Sameshima}, {Schulze}, {Shirakata}, {Silverman}, {Sugiyama},
  {Tait}, {Takada}, {Takata}, {Tanaka}, {Tang}, {Toba}, {Utsumi}, {Wang}, \&
  {Yamashita}}]{matsuoka2018}
{Matsuoka}, Y., {et~al.} 2018, \apjs, 237, 5

\bibitem[{{Mazzucchelli} {et~al.}(2017){Mazzucchelli}, {Ba{\~n}ados},
  {Venemans}, {Decarli}, {Farina}, {Walter}, {Eilers}, {Rix}, {Simcoe},
  {Stern}, {Fan}, {Schlafly}, {De Rosa}, {Hennawi}, {Chambers}, {Greiner},
  {Burgett}, {Draper}, {Kaiser}, {Kudritzki}, {Magnier}, {Metcalfe}, {Waters},
  \& {Wainscoat}}]{mazzucchelli2017}
{Mazzucchelli}, C., {et~al.} 2017, \apj, 849, 91

\bibitem[{{Mu{\~n}oz} \& {Oh}(2016)}]{munoz.oh2016}
{Mu{\~n}oz}, J.~A., \& {Oh}, S.~P. 2016, \mnras, 463, 2085

\bibitem[{{Nagao} {et~al.}(2011){Nagao}, {Maiolino}, {Marconi}, \&
  {Matsuhara}}]{nagao2011}
{Nagao}, T., {Maiolino}, R., {Marconi}, A., \& {Matsuhara}, H. 2011, \aap, 526,
  A149

\bibitem[{{Narayanan} \& {Krumholz}(2017)}]{narayanan.krumholz2017}
{Narayanan}, D., \& {Krumholz}, M.~R. 2017, \mnras, 467, 50

\bibitem[{{Pereira-Santaella} {et~al.}(2017){Pereira-Santaella}, {Rigopoulou},
  {Farrah}, {Lebouteiller}, \& {Li}}]{pereira-sataella2017}
{Pereira-Santaella}, M., {Rigopoulou}, D., {Farrah}, D., {Lebouteiller}, V., \&
  {Li}, J. 2017, \mnras, 470, 1218

\bibitem[{{Priddey} \& {McMahon}(2001)}]{Priddey.McMahon2001}
{Priddey}, R.~S., \& {McMahon}, R.~G. 2001, \mnras, 324, L17

\bibitem[{{Rybak} {et~al.}(2019){Rybak}, {Calistro Rivera}, {Hodge}, {Smail},
  {Walter}, {van der Werf}, {da Cunha}, {Chen}, {Dannerbauer}, {Ivison},
  {Karim}, {Simpson}, {Swinbank}, \& {Wardlow}}]{rybak2019}
{Rybak}, M., {et~al.} 2019, \apj, 876, 112

\bibitem[{{Schneider} {et~al.}(2015){Schneider}, {Bianchi}, {Valiante},
  {Risaliti}, \& {Salvadori}}]{schneider2015}
{Schneider}, R., {Bianchi}, S., {Valiante}, R., {Risaliti}, G., \& {Salvadori},
  S. 2015, \aap, 579, A60

\bibitem[{{Shao} {et~al.}(2019){Shao}, {Wang}, {Carilli}, {Wagg}, {Walter},
  {Li}, {Fan}, {Jiang}, {Riechers}, {Bertoldi}, {Strauss}, {Cox}, {Omont}, \&
  {Menten}}]{shao2019}
{Shao}, Y., {et~al.} 2019, \apj, 876, 99

\bibitem[{{Spinoglio} {et~al.}(2015){Spinoglio}, {Pereira-Santaella}, {Dasyra},
  {Calzoletti}, {Malkan}, {Tommasin}, \& {Busquet}}]{spinoglio2015}
{Spinoglio}, L., {Pereira-Santaella}, M., {Dasyra}, K.~M., {Calzoletti}, L.,
  {Malkan}, M.~A., {Tommasin}, S., \& {Busquet}, G. 2015, \apj, 799, 21

\bibitem[{{Stacey} {et~al.}(2010){Stacey}, {Hailey-Dunsheath}, {Ferkinhoff},
  {Nikola}, {Parshley}, {Benford}, {Staguhn}, \& {Fiolet}}]{stacey2010}
{Stacey}, G.~J., {Hailey-Dunsheath}, S., {Ferkinhoff}, C., {Nikola}, T.,
  {Parshley}, S.~C., {Benford}, D.~J., {Staguhn}, J.~G., \& {Fiolet}, N. 2010,
  \apj, 724, 957

\bibitem[{{Symeonidis}(2017)}]{symeonidis2017}
{Symeonidis}, M. 2017, \mnras, 465, 1401

\bibitem[{{Symeonidis} {et~al.}(2016){Symeonidis}, {Giblin}, {Page}, {Pearson},
  {Bendo}, {Seymour}, \& {Oliver}}]{symeonidis2016}
{Symeonidis}, M., {Giblin}, B.~M., {Page}, M.~J., {Pearson}, C., {Bendo}, G.,
  {Seymour}, N., \& {Oliver}, S.~J. 2016, \mnras, 459, 257

\bibitem[{{Tamura} {et~al.}(2019){Tamura}, {Mawatari}, {Hashimoto}, {Inoue},
  {Zackrisson}, {Christensen}, {Binggeli}, {Matsuda}, {Matsuo}, {Takeuchi},
  {Asano}, {Sunaga}, {Shimizu}, {Okamoto}, {Yoshida}, {Lee}, {Shibuya},
  {Taniguchi}, {Umehata}, {Hatsukade}, {Kohno}, \& {Ota}}]{tamura2019}
{Tamura}, Y., {et~al.} 2019, \apj, 874, 27

\bibitem[{{Valiante} {et~al.}(2017){Valiante}, {Agarwal}, {Habouzit}, \&
  {Pezzulli}}]{valiante2017}
{Valiante}, R., {Agarwal}, B., {Habouzit}, M., \& {Pezzulli}, E. 2017, \pasa,
  34, e031

\bibitem[{{Venemans} {et~al.}(2016){Venemans}, {Walter}, {Zschaechner},
  {Decarli}, {De Rosa}, {Findlay}, {McMahon}, \& {Sutherland}}]{venemans2016}
{Venemans}, B.~P., {Walter}, F., {Zschaechner}, L., {Decarli}, R., {De Rosa},
  G., {Findlay}, J.~R., {McMahon}, R.~G., \& {Sutherland}, W.~J. 2016, \apj,
  816, 37

\bibitem[{{Venemans} {et~al.}(2012){Venemans}, {McMahon}, {Walter}, {Decarli},
  {Cox}, {Neri}, {Hewett}, {Mortlock}, {Simpson}, \& {Warren}}]{venemans2012}
{Venemans}, B.~P., {et~al.} 2012, \apjl, 751, L25

\bibitem[{{Venemans} {et~al.}(2017{\natexlab{a}}){Venemans}, {Walter},
  {Decarli}, {Ba{\~n}ados}, {Carilli}, {Winters}, {Schuster}, {da Cunha},
  {Fan}, {Farina}, {Mazzucchelli}, {Rix}, \& {Weiss}}]{venemans2017.z7.5}
{Venemans}, B.~P., {et~al.} 2017{\natexlab{a}}, \apjl, 851, L8

\bibitem[{{Venemans} {et~al.}(2017{\natexlab{b}}){Venemans}, {Walter},
  {Decarli}, {Ferkinhoff}, {Wei{\ss}}, {Findlay}, {McMahon}, {Sutherland}, \&
  {Meijerink}}]{venemans2017.co}
{Venemans}, B.~P., {et~al.} 2017{\natexlab{b}}, \apj, 845, 154

\bibitem[{{Venemans} {et~al.}(2018){Venemans}, {Decarli}, {Walter},
  {Ba{\~n}ados}, {Bertoldi}, {Fan}, {Farina}, {Mazzucchelli}, {Riechers},
  {Rix}, {Wang}, \& {Yang}}]{venemans2018}
{Venemans}, B.~P., {et~al.} 2018, \apj, 866, 159

\bibitem[{{Walter} {et~al.}(2018){Walter}, {Riechers}, {Novak}, {Decarli},
  {Ferkinhoff}, {Venemans}, {Ba{\~n}ados}, {Bertoldi}, {Carilli}, {Fan},
  {Farina}, {Mazzucchelli}, {Neeleman}, {Rix}, {Strauss}, {Uzgil}, \&
  {Wang}}]{walter2018}
{Walter}, F., {et~al.} 2018, \apjl, 869, L22

\bibitem[{{Wang} {et~al.}(2008){Wang}, {Carilli}, {Wagg}, {Bertoldi}, {Walter},
  {Menten}, {Omont}, {Cox}, {Strauss}, {Fan}, {Jiang}, \&
  {Schneider}}]{wang2008}
{Wang}, R., {et~al.} 2008, \apj, 687, 848

\bibitem[{{Wang} {et~al.}(2010){Wang}, {Carilli}, {Neri}, {Riechers}, {Wagg},
  {Walter}, {Bertoldi}, {Menten}, {Omont}, {Cox}, \& {Fan}}]{wang2010}
{Wang}, R., {et~al.} 2010, \apj, 714, 699

\bibitem[{{Wang} {et~al.}(2013){Wang}, {Wagg}, {Carilli}, {Walter}, {Lentati},
  {Fan}, {Riechers}, {Bertoldi}, {Narayanan}, {Strauss}, {Cox}, {Omont},
  {Menten}, {Knudsen}, {Neri}, \& {Jiang}}]{wang2013}
{Wang}, R., {et~al.} 2013, \apj, 773, 44

\bibitem[{{Wang} {et~al.}(2016){Wang}, {Wu}, {Neri}, {Fan}, {Walter},
  {Carilli}, {Momjian}, {Bertoldi}, {Strauss}, {Li}, {Wang}, {Riechers},
  {Jiang}, {Omont}, {Wagg}, \& {Cox}}]{wang2016}
{Wang}, R., {et~al.} 2016, \apj, 830, 53

\bibitem[{{Wang} {et~al.}(2019){Wang}, {Shao}, {Carilli}, {Jones}, {Walter},
  {Fan}, {Riechers}, {Decarli}, {Bertoldi}, {Wagg}, {Strauss}, {Omont}, {Cox},
  {Jiang}, {Narayanan}, {Menten}, \& {Venemans}}]{wang2019}
{Wang}, R., {et~al.} 2019, (arXiv:1904.07749)

\bibitem[{{Willott} {et~al.}(2015){Willott}, {Bergeron}, \&
  {Omont}}]{willott2015.qso}
{Willott}, C.~J., {Bergeron}, J., \& {Omont}, A. 2015, \apj, 801, 123

\bibitem[{{Willott} {et~al.}(2013){Willott}, {Omont}, \&
  {Bergeron}}]{willott2013}
{Willott}, C.~J., {Omont}, A., \& {Bergeron}, J. 2013, \apj, 770, 13

\bibitem[{{Wu} {et~al.}(2015){Wu}, {Wang}, {Fan}, {Yi}, {Zuo}, {Bian}, {Jiang},
  {McGreer}, {Wang}, {Yang}, {Yang}, {Thompson}, \& {Beletsky}}]{wu2015}
{Wu}, X.-B., {et~al.} 2015, \nat, 518, 512

\bibitem[{{Zhang} {et~al.}(2018){Zhang}, {Ivison}, {George}, {Zhao}, {Dunne},
  {Herrera-Camus}, {Lewis}, {Liu}, {Naylor}, {Oteo}, {Riechers}, {Smail},
  {Yang}, {Eales}, {Hopwood}, {Maddox}, {Omont}, \& {van der Werf}}]{zhang2018}
{Zhang}, Z.-Y., {et~al.} 2018, \mnras, 481, 59

\end{thebibliography}

\enddocument